# Synthesis of rock-salt MeO–ZnO solid solutions (Me = $Ni^{2+}$, $Co^{2+}$, $Fe^{2+}$, $Mn^{2+}$) at high pressure and high temperature

A.N. Baranov[1], P.S. Sokolov[2,3], O.O. Kurakevych[3], V.A. Tafeenko[1], D. Trots[4], V.L. Solozhenko[3*]

[1] *Chemistry Department, Moscow State University, 119992 Moscow, Russia*

[2] *Materials Science Department, Moscow State University, 119992 Moscow, Russia*

[3] *LSPMTM-CNRS, Université Paris Nord, 93430 Villetaneuse, France*

[4] *HASYLAB-DESY, 22603 Hamburg, Germany*

### Abstract

*Series of metastable $Me_{1-x}Zn_xO$ solid solutions (Me = $Ni^{2+}$, $Co^{2+}$, $Fe^{2+}$, $Mn^{2+}$) with the rock-salt (rs) crystal structure have been synthesized from the binary oxides by quenching from 7.7 GPa and 1450-1650 K. Phase composition of the samples, as well as structural properties and stoichiometry of synthesized solid solutions have been studied by X-ray powder diffraction, both conventional and with synchrotron radiation. The widest ($0.3 \leq x \leq 0.8$) composition range of the existence of individual rock-salt solid solution has been established for the NiO-ZnO system. The bulk rs-$Co_{1-x}Zn_xO$, rs-$Fe_{1-x}Zn_xO$ and rs-$Mn_{1-x}Zn_xO$ solid solutions may be quenched down to ambient conditions only with twice lower ZnO content, i.e. $x \leq 0.5$, 0.5 and 0.4, respectively; while formation of rock-salt solid solutions in the CdO-ZnO system has not been observed in the whole concentration range.*

**Introduction**

Zinc oxide belongs to the family of wide-band-gap semiconductors with strong ionic character of chemical bonds. At ambient conditions ZnO has a hexagonal wurtzite structure ($P6_3mc$) that transforms into rock-salt structure ($Fm3m$) at high pressures [1]. At 300 K the wurzite-to-rock-salt phase transition shows a substantial hysteresis (~9 GPa on compression and ~2 GPa on decompression), while at temperatures about 1000 K the transition pressure decreases down to 6 GPa, and hysteresis tends to zero [2].

The thermodynamic studies of the NiO-ZnO, MnO-ZnO, CoO-ZnO [3] and FeO-ZnO [4] systems has shown that at ambient pressure the equilibrium MeO-ZnO phase diagrams are of eutectic type with a wide region of coexistence of rs-MeO-based and wurtzite ZnO-based solid solutions [3, 4].

Previously we have studied the MgO-ZnO system at high pressures and temperatures by quenching [5] and found that the rs-$Mg_{1-x}Zn_xO$ solid solution can be stabilized over the wide (up to $x = 0.7$) concentration range. Later, the use of X-ray diffraction with synchrotron radiation allowed us to *in situ* study the chemical reactions and phase transitions in the MgO-ZnO system at high pressures and temperatures [6]. As a result, the MgO-ZnO phase diagram at 4.4 GPa has been constructed. Other MeO-ZnO systems have not been systematically studied at high pressures so far.

Recently, it has been shown that the pressure of the wurtzite-to-rock-salt transformation decreases for Co- and Mn-doped ZnO as compared to pure ZnO [7-9]. The high pressure – high temperature (HPHT) synthesis of ZnO-rich rs solid solutions $Mn_{0.25}Zn_{0.75}O$ and $Co_{0.27}Zn_{0.73}O$ in the form of thin films has been reported [8, 9], however, in the case of bulk samples the situation may be different.

Here we report the synthesis of the metastable bulk rock-salt $Me_{1-x}Zn_xO$ solid solutions (Me = $Ni^{2+}$, $Co^{2+}$, $Fe^{2+}$, $Mn^{2+}$) under high pressure and high temperature.

**Experimental**

As starting materials we have used the powders of ZnO (99.99%, Alfa Aesar), CoO (99.99%, Aldrich), MnO (99.99%, Aldrich), NiO (99%, Aldrich), FeO (99%, Aldrich) and CdO (99%, Aldrich). The MeO-ZnO mixtures of various stoichiometries (the molar fraction of ZnO from 0 to 1 with 0.1 concentration step) were thoroughly ground in a mortar, then pressed into pellets and placed into gold capsules. Quenching experiments at 7.7 GPa and 1450-1650 K have been performed using a toroid-type high-pressure apparatus [10] at LPMTM-CNRS. Sample pressure as a function of hydraulic oil pressure was calibrated using room-temperature phase transitions in Bi ($Bi_{I-II}$ at 2.55 GPa and $Bi_{III-V}$ at 7.7 GPa), PbSe (4.2 GPa) and PbTe (5.2 GPa). Temperature calibration was performed using Pt/Pt-10%Rh and chromel-alumel thermocouples without correction for the pressure effect on the thermocouple emf, as well as by using the melting of In (670 K), Sb (950 K), Zn (985 K), Si (1230 K), Al (1430 K), Cu (1670 K), Ni (1970 K) and Fe (2070 K) at 7.7 GPa as temperature reference points.

Samples were gradually compressed to 7.7 GPa at ambient temperature, and then the temperature was continuously increased with a heating rate of about 100 K/min. Duration of heating at required temperature was 10-15 minutes. Then samples were quenched by switching off the power and slowly decompressed.

Phase composition of recovered samples has been studied by X-ray powder diffraction using MZIII Seifert (Cu K$\alpha$ radiation) and G3000 TEXT Inel (Cu K$\alpha$1 radiation) X-ray diffractometers. Precise lattice parameters of synthesized solid solutions have been established using X-ray powder diffraction with synchrotron radiation ($\lambda = 0.49342$ Å) at B2 beamline, HASYLAB-DESY. The FullProf program [11] has been used for Rietveld analysis; the details of data processing are described elsewhere [12].

**Results and discussion**

The synthesis of rs-Me$_{1-x}$Zn$_x$O solid solutions has been performed at highest pressure achievable in our toroid-type apparatus, i.e. 7.7 GPa, and at temperatures in the 1450-1650 K range, in which the bulk diffusion in ZnO is intensive enough for the formation of solid solutions [6].

Figure 1 shows the characteristic powder diffraction pattern of a rock-salt NiO-ZnO solid solution synthesized by quenching from 7.7 GPa and 1470 K. It was found that the lattice parameters of rs-Ni$_{1-x}$Zn$_x$O perfectly follow the linear concentration dependence (Vegard's law) from $a_{ZnO}$ = 4.280 Å to $a_{NiO}$ = 4.176 Å in the whole concentration range of existence of the rock-salt solid solutions (x = 0.3-0.8) (see Inset in Figure 1). Samples with ZnO concentration higher than 0.8 were two-phase mixtures of the NiO-ZnO solid solutions with wurtzite and rock-salt structure. Since at ambient conditions the two-phase region of the NiO-ZnO system has been observed already at x > 0.35 [3], one can conclude that at high pressures and temperatures the ZnO solubility limit in this system shifts toward the ZnO isopleth.

In contrast to rs-Ni$_{1-x}$Zn$_x$O, the bulk rs-Co$_{1-x}$Zn$_x$O, rs-Fe$_{1-x}$Zn$_x$O and rs-Mn$_{1-x}$Zn$_x$O solid solutions may be quenched down to ambient conditions only with twice lower ZnO content, i.e. x ≤ 0.5, 0.5 and 0.4, respectively. In the case of the CdO-ZnO system, no formation of rock-salt solid solutions has been observed in all concentration range under study. The obtained results on lattice parameters and limiting solubilities are summarized in Table 1. Similarly to rs-Ni$_{1-x}$Zn$_x$O, all synthesized rock-salt MeO-ZnO solid solutions well follow Vegard's law in the whole concentration range of existence; e.g. for the rock-salt phases with highest ZnO content the lattice parameters are $a$ = 4.275(1) Å (rs-Co$_{0.5}$Zn$_{0.5}$O), $a$ = 4.3034(1) Å (Fe$_{0.5}$Zn$_{0.5}$O), and $a$ = 4.3647(1) Å (Mn$_{0.6}$Zn$_{0.4}$O).

Analysis of the solubility limits of ZnO in rs-Me$_{1-x}$Zn$_x$O solid solutions (Me = Ni$^{2+}$, Co$^{2+}$, Fe$^{2+}$, Mn$^{2+}$) synthesized at ambient pressure and 1070°K [3] has revealed that the higher ZnO solubility limit is observed for the Me$^{2+}$ cations with lower radius (see Table 1). By analogy, the large extension of the ZnO solubility limits at high pressures and temperatures may be attributed to the decrease of the difference between $Me^{2+}$ and $Zn^{2+}$ cation size ($r_{Me2+}$ - $r_{Zn2+}$). So, the fact that the rs-$Cd_{1-x}Zn_xO$ solid solutions, if exist at HTHP, cannot be quenched down to ambient conditions is probably due to the large cation radius of $Cd^{2+}$ as compared to $Zn^{2+}$.

The estimation for the energetic preference $\Delta H_{6/4}$ of octahedral coordination as compared to tetrahedral one for some bivalent cations [14] is also presented in Table 1. One can see that ZnO solubility limits strongly correlate with $\Delta H_{6/4}$. As it follows from our results, the highest concentration of ZnO (x = 0.8) in the metastable rs-$Me_{1-x}Zn_xO$ solid solutions quenched down to ambient conditions has been observed for Ni, which is in excellent agreement with the fact that $Ni^{2+}$ cation has the highest preference to 6-fold coordination.

**Conclusions**

Thus, we have found that bulk single-phase rock-salt $Me_{1-x}Zn_xO$ solid solutions (Me = $Ni^{2+}$, $Co^{2+}$, $Fe^{2+}$, $Mn^{2+}$) with high ZnO concentration can be synthesized at high pressures and temperatures. The widest ($0.3 \leq x \leq 0.8$) composition range of the existence of individual rock-salt solid solution has been established for the NiO-ZnO system. The rs-$Co_{1-x}Zn_xO$, rs-$Fe_{1-x}Zn_xO$ and rs-$Mn_{1-x}Zn_xO$ solid solutions may be quenched down to ambient conditions only with twice lower ZnO content. Formation of rock-salt solid solutions in the CdO-ZnO system has not been observed.

**Acknowledgements**

X-ray diffraction measurements with synchrotron radiation were carried out during beamtime allocated to Project I-20070033 EC at HASYLAB-DESY. This work has been partially supported by the Russian Foundation for Basic Research (Project № 06-03-33042). PSS is grateful to the French Ministry of Foreign Affairs for funding provided (BGF fellowship No. 2007 1572).

.

**Table 1**

Me$^{2+}$ cation properties (Me$^{2+}$ = Ni$^{2+}$, Mg$^{2+}$, Co$^{2+}$, Fe$^{2+}$, Mn$^{2+}$, Cd$^{2+}$), MeO structural data and ZnO solubility limits in rs-Me$_{1-x}$Zn$_x$O solid solutions synthesized at different pressures and temperatures.

|  | NiO | MgO | **ZnO** | CoO | FeO | MnO | CdO |
|---|---|---|---|---|---|---|---|
| Lattice parameter, Å | 4.176 | 4.213 | **4.280** | 4.258 | 4.326 | 4.448 | 4.689 |
| Cation radius, pm** | 83.0 | 86.0 | **88.0** | 88.5* | 92.0* | 97.0* | 109.0 |
|  |  |  |  | 68.5 | 75.0 | 81.0 |  |
| Electronegativity of cation** | 1.7 | 1.25 | **1.6** | 1.65 | 1.6 | 1.45 | 1.5 |
| $\Delta H_{6/4}$, kcal/mol† | -13.7 | -7.1 | **+5±1** | -7 | -6.3 | -3 | -5 |
| $x_{ZnO}$ in rs-Me$_{1-x}$Zn$_x$O, 0.1 MPa 1070 K‡ | 0.35 | 0.33 | — | 0.22 | 0.20 | 0.09 | — |
| $x_{ZnO}$ in rs-Me$_{1-x}$Zn$_x$O, 7.7 GPa 1550 K | 0.80 | 0.70†† | — | 0.50 | 0.50 | 0.40 | — |

* Radius of cation in a high spin state (bottom number is a radius of cation in low spin state).

† According to Ref. 14.

‡ According to Refs. 3, 4.

** According to Ref. 13.

†† According to Ref. 12.

**Figure 1**

The observed and calculated X-ray diffraction patterns of the rock-salt $Ni_{0.2}Zn_{0.8}O$ solid solution quenched from 7.7 GPa and 1470 K. (Insert: Concentration dependence of the lattice parameter of rs-$Ni_{1-x}Zn_xO$ solid solutions).

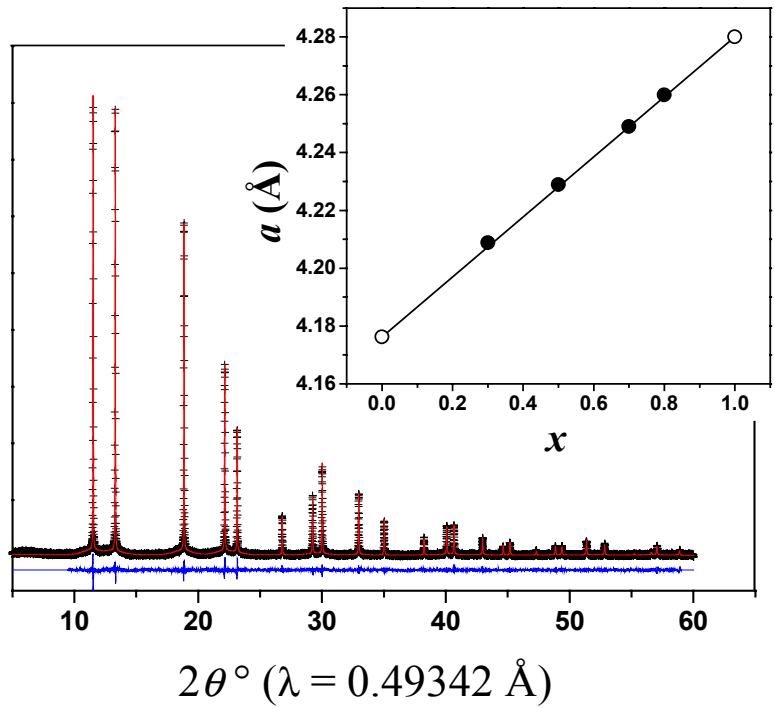